\documentclass[aps,nofootinbib,twocolumn,reprint,superscriptaddress]{revtex4-2}
\usepackage[T1]{fontenc}
\usepackage[utf8]{inputenc}
\usepackage{xcolor}
\usepackage{amsmath}
\usepackage{amssymb}
\usepackage{graphicx}
\usepackage{wasysym}
\usepackage[bookmarks=false,
 breaklinks=false,pdfborder={0 0 1},backref=false,colorlinks=false]
 {hyperref}
\hypersetup{
 linkcolor=blue,anchorcolor=blue,citecolor=blue,urlcolor=blue}

\makeatletter

\newcommand{\noun}[1]{\textsc{#1}}

\usepackage{amsfonts}\usepackage{wasysym}
\usepackage{braket}
\usepackage{appendix}
\allowdisplaybreaks[4]
\hypersetup{
  colorlinks=true,
  linkcolor=blue,    
  citecolor=blue,    
  urlcolor=blue,     
}

\makeatother

\begin{document}
\title{A Mean-Field Approach to the Dielectric Response of Bulk Superconductors
for Light Dark Matter Direct Detection}
\author{Chengyao Gan}
\affiliation{College of Mathematics and Physics, Beijing University of Chemical
Technology, Beijing 100029, China}
\author{Hui Li}
\email{physicslihui@163.com}

\affiliation{Department of Physics, Wuhan University of Technology, Wuhan 430070,
China}
\author{Wenjing Li}
\affiliation{Centre for Quantum Physics, Key Laboratory of Advanced Optoelectronic
Quantum Architecture and Measurement (MOE), School of Physics, Beijing
Institute of Technology, Beijing 100081, China}
\author{Zheng-Liang Liang}
\email{liangzl@mail.buct.edu.cn}

\affiliation{College of Mathematics and Physics, Beijing University of Chemical
Technology, Beijing 100029, China}
\author{Lin Zhang}
\email{zhanglin57@mail.sysu.edu.cn}

\affiliation{School of Science, Shenzhen Campus of Sun Yat-sen University, Shenzhen
518107, China}
\affiliation{Sun Yat-sen University, Guangzhou 510275, China}
\author{Fawei Zheng}
\email{fwzheng@bit.edu.cn}

\affiliation{Centre for Quantum Physics, Key Laboratory of Advanced Optoelectronic
Quantum Architecture and Measurement (MOE), School of Physics, Beijing
Institute of Technology, Beijing 100081, China}
\begin{abstract}
The dielectric function is central to describing many-body screening
effects in dark matter (DM) direct detection with condensed matter
targets. Current superconducting detector analyses employ the free-electron
Lindhard dielectric function to model in-medium effects, an approximation
whose validity in the superconducting state remains untested. We derive
the electronic dielectric function for bulk superconductors within
the Bardeen-Cooper-Schrieffer (BCS) framework, incorporating the full
Bogoliubov quasiparticle coherence factors in the random-phase approximation.
A systematic comparison with the Lindhard function for aluminum and
tungsten silicide (WSi) reveals good agreement for energy depositions
$\omega\gtrsim5\Delta$, establishing the Lindhard function as a robust
approximation for superconducting DM detectors operating in this regime.
\end{abstract}
\maketitle

\section{\noun{Introduction}}

The past decade has seen the rapid maturation of dark matter (DM)
interactions with condensed matter systems as a primary avenue for
sub-GeV DM detection, driving enormous progress across both theoretical
frameworks~\citep{Essig:2011nj,Graham:2012su,Hochberg:2015fth,Essig:2015cda,Griffin:2018bjn,Ibe2018,Knapen2018,Liang:2018bdb,Campbell-Deem:2019hdx,Geilhufe:2019ndy,Griffin:2019mvc,Liang:2019nnx,Trickle:2019nya,Kahn:2020fef,Knapen:2020aky,Kozaczuk:2020uzb,Kurinsky:2020dpb,Trickle:2020oki,Griffin:2021znd,Vahsen:2021gnb,Catena:2021qsr,Hochberg:2021pkt,Knapen:2021run,Wang:2021oha,Knapen:2021bwg,Liang:2022xbu,Berghaus:2022pbu,Campbell-Deem:2022fqm,Li:2022acp,Esposito:2022bnu,Dreyer:2023ovn,Liang:2024xcx,Essig:2024ebk,Dent:2025drd,Liang:2025nan,Esposito:2025iry,Mai:2025zau,Sun:2025gyj}
and experimental implementations~\citep{Amaral:2020ryn,CRESST:2019jnq,Arnaud:2020svb,Barak:2020fql,Castello-Mor:2020jhd,COSINE-100:2021poy,SuperCDMS:2022kgp,CDEX:2022kcd,CONNIE:2024off}
(for a recent review, see Reference \citep{Kahn:2021ttr}). Among
these efforts, detection experiments based on superconducting sensors~\citep{Hochberg:2015pha,Hochberg:2021ymx,Hochberg:2021yud,Gao:2024irf,Griffin:2024jec,Das:2024jdz,QROCODILE:2024nqm,Chen:2025cvl,Schwemmbauer:2025evp}
(e.g., \emph{kinetic inductance detectors} (KIDs)~\citep{Gao:2024irf},
\emph{transition-edge sensors} (TESs)~\citep{Chen:2025cvl,Schwemmbauer:2025evp},
and \emph{superconducting nanowire single-photon detectors} (SNSPDs)~\citep{Hochberg:2021yud,QROCODILE:2024nqm})
are garnering increasing interest. The recent QROCODILE experiment~\citep{QROCODILE:2024nqm},
for instance, based on a microwire SNSPD, achieved a threshold of
0.11\,eV, and placed world\nobreakdash-leading constraints on DM\nobreakdash-electron
scattering, DM\nobreakdash-nucleon scattering for DM masses as low
as 30\,keV. These results demonstrate the power of SNSPDs for probing
light DM parameter space, and future upgrades, including larger arrays
and even lower thresholds, promise to further extend the reach.

As DM cannot be actively manipulated like photons, probing it with
superconducting sensors demands an accurate description of its interactions
with the detector, particularly the many-body electronic effects within
the superconductors. In the context of the linear response theory,
the rate for DM-electron scattering is governed by the dynamic structure
factor, which is related to the imaginary part of the inverse dielectric
function. For normal metals, the Lindhard dielectric function is widely
used to give a description of the electronic response within the \emph{random\nobreakdash-phase
approximation} (RPA). The Lindhard function was also employed in the
superconductor-based experiments to model the screening effect~\citep{Hochberg:2021ymx,Hochberg:2021yud,QROCODILE:2024nqm}.
This approach relies on the assumption that the normal-state dielectric
response adequately approximates the screening in the superconducting
state. While this approximation is computationally convenient and
has been widely used, its validity has not been rigorously tested.

In this work, we generalize the RPA framework to the superconducting
ground state in the \emph{Bardeen, Cooper, and Schrieffer} (BCS) theory~\citep{Bardeen:1957mv}.
By deriving the quasiparticle polarization function that includes
the full coherence factors, the corresponding dielectric function
$\epsilon_{\mathrm{BCS}}\left(\mathbf{Q},\omega\right)$ is obtained
within the superconducting phase. We then perform a systematic comparison
between $\epsilon_{\mathrm{BCS}}\left(\mathbf{Q},\omega\right)$ and
the Lindhard function $\epsilon_{\mathrm{Lin}}\left(\mathbf{Q},\omega\right)$
for common materials relevant for SNSPD experiment, specifically aluminum
and tungsten silicide (WSi), the latter being the active material
in the QROCODILE detector. Natural units ($\hbar=c=1$) are used throughout
the paper.

\section{\textit{\emph{Quasiparticles in Nambu Formalism}}}

In the Nambu spinor formulation, the electron field is organized as
a doublet $\psi_{\mathbf{p}}=(c_{\mathbf{p},\uparrow},c^{\dagger}_{-\mathbf{p},\downarrow})^{T}$,
a representation that naturally groups electron and hole degrees of
freedom. The Coulomb interaction between electrons, which plays a
central role in the screening physics we investigate, is most conveniently
expressed in this basis. In momentum space, the electron density operator
transforms as $\sum_{\mathbf{p},\sigma}c^{\dagger}_{\mathbf{p},\sigma}c_{\mathbf{p}+\mathbf{k},\sigma}=\sum_{\mathbf{p}}\psi^{\dagger}_{\mathbf{p}-\mathbf{k}}\tau^{3}\psi_{\mathbf{p}}$,
where $\tau^{3}=\operatorname{diag}(1,-1)$ is the Pauli matrix in
isospin space, its diagonal elements reflecting the opposite charge
of electrons ($+1$) and holes ($-1$). The full Coulomb interaction
Lagrangian then takes the compact form (see the Appendix for a step-by-step
derivation)
\begin{equation}
\begin{aligned}[b]\int\mbox{d}^{4}x\,\mathcal{L}_{ee} & =-\frac{1}{2V}\sum_{\mathbf{p},\mathbf{p}',\mathbf{k}}\int\mbox{d}t\,\psi^{\dagger}_{\mathbf{p}-\mathbf{k}}\tau^{3}\psi_{\mathbf{p}}\\
 & \quad\times V^{\mathrm{C}}(-\mathbf{k})\,\psi^{\dagger}_{\mathbf{p}'+\mathbf{k}}\tau^{3}\psi_{\mathbf{p}'},
\end{aligned}
\label{eq:Coulomb_Nambu}
\end{equation}
where $V^{\mathrm{C}}(-\mathbf{k})=4\pi\alpha/k^{2}$ ($k=|\mathbf{k}|$)
is the momentum-space Coulomb potential, with $\alpha=e^{2}/(4\pi)$
being the electromagnetic fine-structure constant.

In the superconducting phase, the BCS mean-field Hamiltonian introduces
pairing correlations that couple electron and hole sectors. The electron
Lagrangian reads
\begin{equation}
\int\mbox{d}^{4}x\,\mathcal{L}_{e}=\int\mbox{d}t\sum_{\mathbf{p}}\psi^{\dagger}_{\mathbf{p}}\!\left[i\frac{\partial}{\partial t}-\begin{pmatrix}\varepsilon_{\mathbf{p}} & \Delta\\
\Delta & -\varepsilon_{\mathbf{p}}
\end{pmatrix}\right]\!\psi_{\mathbf{p}},
\end{equation}
where $\varepsilon_{\mathbf{p}}=|\mathbf{p}|^{2}/(2m_{e})-E_{F}$
is the single-particle energy measured from the Fermi level, and $\Delta$
is the superconducting gap, determined self-consistently from the
BCS gap equation. The off-diagonal pairing field $\Delta$ renders
the electron and hole states no longer eigenstates of the system,
necessitating a change of basis.

The Bogoliubov unitary transformation $U_{\mathbf{p}}$ diagonalizes
the $2\times2$ Hamiltonian matrix:
\begin{equation}
U^{\dagger}_{\mathbf{p}}\begin{pmatrix}\varepsilon_{\mathbf{p}} & \Delta\\
\Delta & -\varepsilon_{\mathbf{p}}
\end{pmatrix}U_{\mathbf{p}}=\operatorname{diag}(E_{\mathbf{p}},-E_{\mathbf{p}}),
\end{equation}
yielding the quasiparticle dispersion $E_{\mathbf{p}}=\sqrt{\varepsilon^{2}_{\mathbf{p}}+\Delta^{2}}$,
which exhibits the characteristic gap $2\Delta$ in the excitation
spectrum. The transformation matrix is parameterized as $U_{\mathbf{p}}=\bigl(\begin{smallmatrix}u_{\mathbf{p}} & -v_{\mathbf{p}}\\
v_{\mathbf{p}} & u_{\mathbf{p}}
\end{smallmatrix}\bigr)$ with the coherence factors $\left(u_{\mathbf{p}},v_{\mathbf{p}}\right)=\left(\sqrt{\frac{1}{2}\left(1+\frac{\varepsilon_{\mathbf{p}}}{E_{\mathbf{p}}}\right)},\sqrt{\frac{1}{2}\left(1-\frac{\varepsilon_{\mathbf{p}}}{E_{\mathbf{p}}}\right)}\right)$.
Physically, $v^{2}_{\mathbf{p}}$ represents the probability that
the pair state $(\mathbf{p}\uparrow,-\mathbf{p}\downarrow)$ is occupied
in the BCS ground state, while $u^{2}_{\mathbf{p}}=1-v^{2}_{\mathbf{p}}$
gives the probability it is empty. The quasiparticle field $\Psi_{\mathbf{p}}=U^{\dagger}_{\mathbf{p}}\psi_{\mathbf{p}}=(\gamma_{\mathbf{p},\uparrow},\gamma^{\dagger}_{-\mathbf{p},\downarrow})^{T}$
diagonalizes the BCS action and constitutes the appropriate eigenbasis
for describing excitations above the superconducting condensate. The
inverse transformation $\gamma_{\mathbf{p},\uparrow}=u_{\mathbf{p}}c_{\mathbf{p},\uparrow}+v_{\mathbf{p}}c^{\dagger}_{-\mathbf{p},\downarrow}$
reveals that QPs are coherent superpositions of electron and hole
states, with propagator
\begin{equation}
S_{\mathbf{p}}(\omega)=\operatorname{diag}\!\left(\frac{1}{\omega-E_{\mathbf{p}}+i0^{+}},\;\frac{1}{\omega+E_{\mathbf{p}}-i0^{+}}\right),\label{eq:QP_propagator}
\end{equation}
where the infinitesimal $i0^{+}$ stems from the QP one-particle spectral
function. (A detailed review of the BCS formalism is provided in the
Appendix.)

\section{\textit{\emph{DM-QP Interaction and BCS dielectric function}}}

For concreteness, in this work we consider a generic model where spin-$1/2$
DM particles ($\chi$) couple to the density of electrons via a Yukawa
potential $V^{\chi e}\left(\mathbf{x}-\mathbf{x}'\right)\propto e^{-m_{A'}\left|\mathbf{x}-\mathbf{x}'\right|}/\left|\mathbf{x}-\mathbf{x}'\right|$,
with $m_{A'}$ being the mass of the mediator $A'$. Similarly to
the above discussion, the relevant action can be also written in terms
of the QP field:
\begin{widetext}
\begin{eqnarray}
\int\mathrm{d}^{4}x\mathcal{L}_{\chi e} & = & -\frac{1}{V}\sum_{\mathbf{p}_{\chi},\sigma_{\chi}}\sum_{\mathbf{p},\mathbf{k}}\int\mathrm{d}t\,a^{\dagger}_{\mathbf{p}_{\chi},\sigma_{\chi}}\left(t\right)a_{\mathbf{\mathbf{p}_{\chi}+\mathbf{k},\sigma_{\chi}}}\left(t\right)\varPsi^{\dagger}_{\mathbf{p}+\mathbf{k}}\left(t\right)U^{\dagger}_{\mathbf{p}+\mathbf{k}}\tau^{3}U_{\mathbf{p}}\varPsi_{\mathbf{p}}\left(t\right)V^{\chi e}\left(-\mathbf{\mathbf{k}}\right),\label{eq:DM-e_lagrangian}
\end{eqnarray}
where $\tau^{3}=\operatorname{diag}(1,-1)$ is the isospin matrix,
$a_{\mathbf{\mathbf{p}_{\chi},\sigma_{\chi}}}$ ($a^{\dagger}_{\mathbf{\mathbf{p}_{\chi},\sigma_{\chi}}}$)
annihilates (creates) a DM particle with momentum $\mathbf{p}_{\chi}$
and spin $\sigma_{\chi}=\uparrow,\downarrow$, and $V^{\chi e}\left(\mathbf{k}\right)$
is the DM-electron interaction in momentum space. If an incident DM
particle deposits energy exceeding the minimum pair-breaking threshold
of $2\Delta$, it breaks a Cooper pair and excites two QPs from the
BCS vacuum $\left|0;\mathrm{BCS}\right\rangle $. The four-fermion
interaction in Eq.~(\ref{eq:DM-e_lagrangian}) directly describes
the scattering process $\left|\mathbf{p}_{\chi}\right\rangle \left|0;\mathrm{BCS}\right\rangle \rightarrow\left|\mathbf{p}_{\chi}'\right\rangle \left|\mathbf{p}_{1},\uparrow;\mathbf{p}_{2},\downarrow\right\rangle _{\mathrm{QP}}$
at tree level, where $\mathbf{p}_{1}$ and $\mathbf{p}_{2}$ represent
the momenta of the two outgoing QPs. It is noted that the simple model
in Eq.~(\ref{eq:DM-e_lagrangian}) does not change the spin of the
DM particle in such a scattering, and thus the spins of the outgoing
QPs are oriented in opposite directions. The tree-level $\mathcal{T}$-matrix
element for $|\mathbf{p}_{\chi}\rangle|0;\mathrm{BCS}\rangle\to|\mathbf{p}'_{\chi}\rangle|\mathbf{p}_{1},\uparrow;\mathbf{p}_{2},\downarrow\rangle_{\mathrm{QP}}$
is
\begin{equation}
\braket{\mathbf{p}_{\chi}';\mathrm{QP}_{1},\uparrow;\mathrm{QP}_{2},\downarrow|\,i\mathcal{T}\,|\mathbf{p}_{\chi};0,\mathrm{BCS}}=i(u_{\mathbf{p}_{1}}v_{\mathbf{p}_{2}}+v_{\mathbf{p}_{1}}u_{\mathbf{p}_{2}})V^{\chi e}(-\mathbf{p}_{1}-\mathbf{p}_{2})\,\frac{2\pi}{V}\,\delta(E_{\mathbf{p}_{1}}+E_{\mathbf{p}_{2}}+\varepsilon_{\mathbf{p}'_{\chi}}-\varepsilon_{\mathbf{p}_{\chi}}),\label{eq:tree_amp}
\end{equation}
where the BCS coherence factor $(u_{\mathbf{p}_{1}}v_{\mathbf{p}_{2}}+v_{\mathbf{p}_{1}}u_{\mathbf{p}_{2}})$
reflects the projection of the electron-density vertex onto the QP
basis. This process and its RPA renormalization are illustrated in
Fig.~\ref{fig:Feynman-diagrams}.

\begin{figure*}[t]
\centering \includegraphics[scale=0.34]{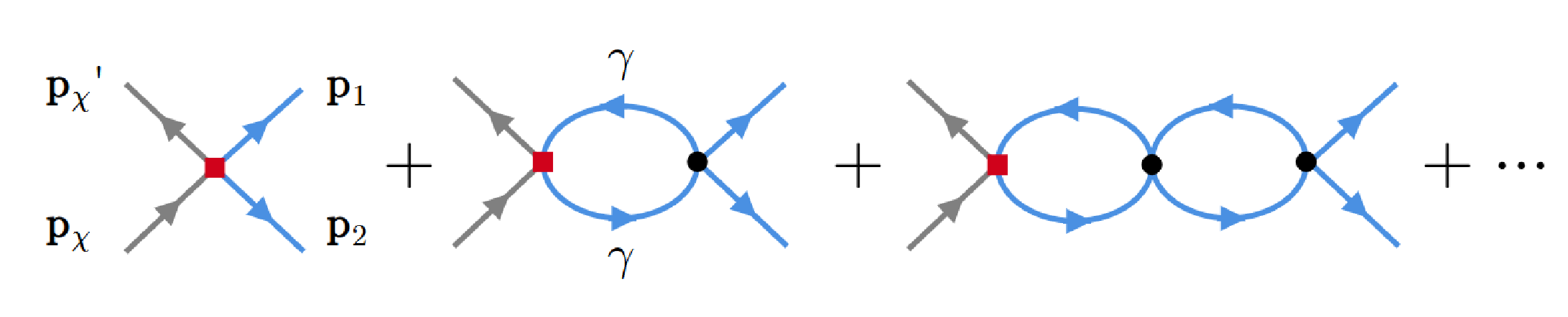} \caption{Feynman diagrams for the scattering process $|\mathbf{p}_{\chi}\rangle|0;\mathrm{BCS}\rangle\to|\mathbf{p}'_{\chi}\rangle|\mathbf{p}_{1};\mathbf{p}_{2}\rangle_{\mathrm{QP}}$
under the RPA. Gray (blue) lines denote DM (QPs), the red box indicates
the DM-electron coupling, and the black circle represents the Coulomb
interaction between QPs.}\label{fig:Feynman-diagrams}
\end{figure*}

Many-body screening effect is incorporated within the RPA by summing
QP bubble diagrams to all orders, as illustrated in Fig.~\ref{fig:Feynman-diagrams}.
The screened amplitude becomes proportional to $V^{\chi e}(\mathbf{Q})/\epsilon_{\mathrm{BCS}}(\mathbf{Q},\omega)$,
where the BCS dielectric function $\epsilon_{\mathrm{BCS}}(\mathbf{Q},\omega)=1-V^{\mathrm{C}}(\mathbf{Q})\,\Pi_{\mathrm{QP}}(\mathbf{Q},\omega)$
is defined in terms of the Coulomb potential $V^{\mathrm{C}}(\mathbf{Q})=4\pi\alpha/Q^{2}$.
The QP polarization function $\Pi_{\mathrm{QP}}$ follows from evaluating
the bubble diagram:
\begin{align}
\Pi_{\mathrm{QP}}(\mathbf{Q},\omega) & =\frac{1}{V}\int\frac{\mbox{d}E}{2\pi i}\sum_{\mathbf{p}}\operatorname{tr}\!\bigl[U^{\dagger}_{\mathbf{p}+\mathbf{Q}}\tau^{3}U_{\mathbf{p}}S_{\mathbf{p}}(E)U^{\dagger}_{\mathbf{p}}\tau^{3}U_{\mathbf{p}+\mathbf{Q}}S_{\mathbf{p}+\mathbf{Q}}(\omega+E)\bigr]\nonumber \\
 & =\frac{1}{2V}\sum_{\mathbf{p}}\!\left(1-\frac{\varepsilon_{\mathbf{p}}\varepsilon_{\mathbf{p}+\mathbf{Q}}-\Delta^{2}}{E_{\mathbf{p}}E_{\mathbf{p}+\mathbf{Q}}}\right)\!\left[\frac{1}{\omega-E_{\mathbf{p}}-E_{\mathbf{p}+\mathbf{Q}}+i0^{+}}-\frac{1}{\omega+E_{\mathbf{p}}+E_{\mathbf{p}+\mathbf{Q}}-i0^{+}}\right]\!.\label{eq:Pi_QP}
\end{align}
Substituting Eq.~(\ref{eq:Pi_QP}) into the expression for $\epsilon_{\mathrm{BCS}}$,
the full BCS electronic dielectric function at zero temperature is
\begin{equation}
\epsilon_{\mathrm{BCS}}(\mathbf{Q},\omega)=1-\frac{2\pi\alpha}{VQ^{2}}\sum_{\mathbf{p}}\!\left(1-\frac{\varepsilon_{\mathbf{p}}\varepsilon_{\mathbf{p}+\mathbf{Q}}-\Delta^{2}}{E_{\mathbf{p}}E_{\mathbf{p}+\mathbf{Q}}}\right)\!\left[\frac{1}{\omega-E_{\mathbf{p}+\mathbf{Q}}-E_{\mathbf{p}}+i0^{+}}-\frac{1}{\omega+E_{\mathbf{p}+\mathbf{Q}}+E_{\mathbf{p}}-i0^{+}}\right]\!.\label{eq:eps_BCS}
\end{equation}
\end{widetext}

The zero-temperature limit is a reasonable approximation, as superconducting
DM detectors operate at millikelvin temperatures. In the limit $\Delta\to0$,
$\epsilon_{\mathrm{BCS}}$ reduces to the Lindhard function for $\omega>0$.

By combining Eq.~(\ref{eq:tree_amp}) and Eq.~(\ref{eq:eps_BCS}),
we obtain the QP production rate for a DM particle incident on a superconductor
with velocity $\mathbf{v}_{\chi}$:

\begin{align}
\Gamma\left(\mathbf{v}_{\chi}\right) & =\int\mathrm{d}\omega\int^{Q^{+}}_{Q^{-}}\frac{Q^{3}\mathrm{d}Q}{\left(2\pi\right)^{3}}\frac{\left|V^{\chi e}\left(\mathbf{Q}\right)\right|^{2}}{\alpha\,v_{\chi}}\nonumber \\
 & \quad\times\frac{\mathrm{Im}\left[\epsilon_{\mathrm{BCS}}\left(\mathbf{Q},\omega\right)\right]}{\left|\epsilon_{\mathrm{BCS}}\left(\mathbf{Q},\omega\right)\right|^{2}},\label{eq:QP_production_rate}
\end{align}
where the integration limits are given by $Q^{\pm}=m_{\chi}v_{\chi}\left(1\pm\sqrt{1-\frac{2\omega}{m_{\chi}v^{2}_{\chi}}}\right)$,
and $v_{\chi}=\left|\mathbf{v}_{\chi}\right|$ is the magnitude of
the DM velocity, $m_{\chi}$ represents the DM particle mass. In momentum
space, the Yukawa potential is expressed as $\left|V^{\chi e}\left(\mathbf{Q}\right)\right|^{2}=\pi\bar{\sigma}_{e}\mu^{-2}_{\chi e}\left(Q^{2}_{0}+m^{2}_{A'}\right)^{2}/\left(Q^{2}+m^{2}_{A'}\right)^{2}$,
where $Q_{0}=\alpha m_{e}$ represents the characteristic momentum
transfer in atomic scattering, $\mu_{\chi e}$ denotes the reduced
mass of the DM-electron system, and $\bar{\sigma}_{e}$ serves as
a reference cross section parameterizing the strength of the DM-electron
coupling. $\mathrm{Im}\left[\epsilon_{\mathrm{BCS}}/\left|\epsilon_{\mathrm{BCS}}\right|^{2}\right]=\mathrm{Im}\left[-1/\epsilon_{\mathrm{BCS}}\right]$
is the\emph{ energy loss function} (ELF) of a superconductor. It is
noted that the dielectric function is independent of the direction
of the momentum $\mathbf{Q}$, \emph{i.e.}, $\epsilon_{\mathrm{BCS}}\left(\mathbf{Q},\omega\right)=\epsilon_{\mathrm{BCS}}\left(Q,\omega\right)$,
as the isotropic symmetry is assumed for the electron system.
\begin{figure*}[t]
\centering \includegraphics[scale=0.75]{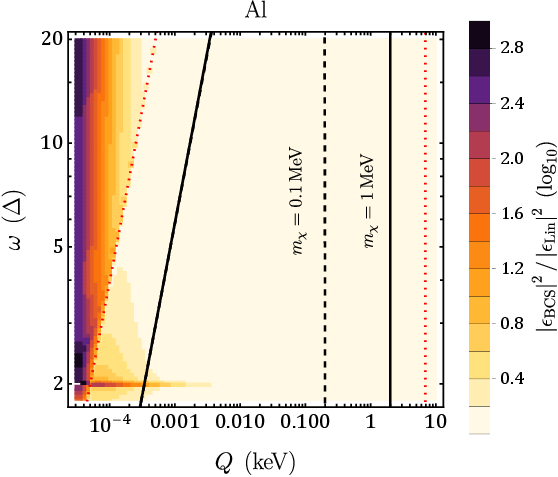}\hspace{1.5cm}\includegraphics[scale=0.75]{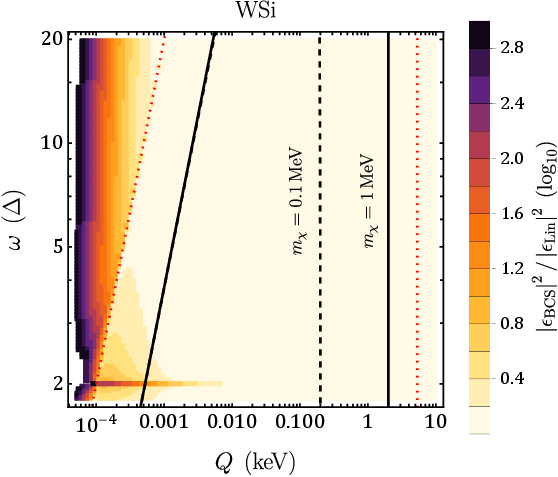}
\caption{Ratio of squared dielectric functions $|\epsilon_{\mathrm{BCS}}|^{2}/|\epsilon_{\mathrm{Lin}}|^{2}$
for aluminum (left) and WSi (right). Energy is in units of $\Delta$.
Red dotted curves indicate the boundaries where $\operatorname{Im}[\epsilon_{\mathrm{Lin}}]$
vanishes. Black solid (dashed) curves show the kinematic contour for
$m_{\chi}=1\,\mathrm{MeV}$ ($0.1\,\mathrm{MeV}$) DM with $v_{\chi}=10^{-3}c$.}
\label{fig:Ratios}
\end{figure*}

\section{\textit{\emph{Numerical Results}}}

In experiments such as QROCODILE~\citep{QROCODILE:2024nqm}, which
employs SNSPDs, the BCS electronic dielectric function $\epsilon_{\mathrm{BCS}}$
is approximated by the Lindhard function $\epsilon_{\mathrm{Lin}}$~,
\emph{i.e}.,
\begin{equation}
\begin{aligned}[b]\Gamma\left(\mathbf{v}_{\chi}\right) & \simeq\int\mathrm{d}\omega\int^{Q^{+}}_{Q^{-}}\frac{Q^{3}\mathrm{d}Q}{\left(2\pi\right)^{3}}\frac{\left|V^{\chi e}\left(\mathbf{Q}\right)\right|^{2}}{\alpha\,v_{\chi}}\\
 & \quad\times\frac{\mathrm{Im}\left[\epsilon_{\mathrm{BCS}}\left(\mathbf{Q},\omega\right)\right]}{\left|\epsilon_{\mathrm{Lin}}\left(\mathbf{Q},\omega\right)\right|^{2}}.
\end{aligned}
\label{eq:approximated_Rate}
\end{equation}
While this approximation retains the BCS coherence factor $\left(u_{\mathrm{\mathbf{p}_{1}}}v_{\mathbf{p}_{2}}+v_{\mathbf{p}_{1}}u_{\mathrm{\mathbf{p}_{2}}}\right)^{2}$
for the tree-level QP production rate, it approximates superconducting
screening with the normal-state dielectric function. Given that the
Lindhard function characterizes the RPA dielectric response of a \emph{homogeneous
electron gas} (HEG) in normal metals, its validity in the superconducting
state has yet to be verified.

The accuracy of this approximation can now be examined. Taking aluminum
and WSi as examples (the material used in the QROCODILE experiment~\citep{QROCODILE:2024nqm}),
the BCS dielectric function $\epsilon_{\mathrm{BCS}}$ is numerically
calculated using an energy gap of $\Delta=0.17\,\mathrm{meV}$, and
a Fermi energy $E_{F}=11.7\,\mathrm{eV}$ for aluminum, and $\Delta=0.26\,\mathrm{meV}$,
$E_{F}=7.0\,\mathrm{eV}$ for WSi, respectively. We present the ratio
$\left|\epsilon_{\mathrm{BCS}}\left(\mathbf{Q},\omega\right)\right|^{2}/\left|\epsilon_{\mathrm{Lin}}\left(\mathbf{Q},\omega\right)\right|^{2}$
in Fig.~\ref{fig:Ratios}, which directly reflects the difference
in the QP production rates calculated from the two dielectric functions.

It is observed that $\left|\epsilon_{\mathrm{BCS}}\right|^{2}$ and
$\left|\epsilon_{\mathrm{Lin}}\right|^{2}$ agree closely over most
of the parameter space of interest, except in two specific regimes:
(1) above the red dotted curve $\omega=\left(\frac{Q}{2}+p_{F}\right)\frac{Q}{m_{e}}\simeq v_{F}Q$
(with $p_{F}$ being the Fermi momentum), where the imaginary part
of the Lindhard function vanishes (see the Appendix) while that of
the electronic BCS dielectric function remains finite; and (2) near
the edge of the energy gap $\omega=2\Delta$, only above which the
QP production is possible.

We also plot the kinematic regimes relevant for incident DM with a
typical halo velocity of $v_{\chi}=10^{-3}c$. The momentum transfer
$Q$ and energy deposition $\omega$ are bounded by $\omega\leq Qv_{\chi}-Q^{2}/\left(2m_{\chi}\right)$,
shown by the contours for two benchmark DM masses, in black solid
($m_{\chi}=1\,\mbox{MeV}$) and dashed ($m_{\chi}=0.1\,\mbox{MeV}$)
lines, respectively. The two curves coincide in the low-$Q$ region,
converging to the asymptotic limit $\omega=v_{\chi}Q$.

Our numerical results confirm that the Lindhard dielectric function
provides an accurate description of the electronic screening effect
in QP production induced by halo DM in bulk superconductors. For experiments
based on superconducting nanowires or thin films with a typical layer
size of $10\,\mbox{nm}$, the geometric finite-size effects of the
superconductor can be considered negligible for momentum transfer
$Q\gtrsim0.2\,\mbox{keV}$ (or equivalently, $1/Q\lesssim1\,\mbox{nm}$).
Consequently, the results presented here are applicable to DM particles
with masses $m_{\chi}\apprge0.1\,\mbox{MeV}$, where the bulk approximation
remains valid. In the case of thin layer size of $\mathcal{O}\left(1\,\mbox{nm}\right)$,
Refs~\citep{Lasenby:2021wsc,QROCODILE:2024nqm} computed the response
function by numerically solving Maxwell’s equations with the appropriate
boundary conditions for free electrons. Given that the free-electron-based
Lindhard function effectively describes the screening effect in bulk
superconductors, it is natural to expect this free-electron-based
numerical method can be extended to superconducting thin films.

\section{\textit{\emph{Conclusion and discussion}}}

In summary, within the RPA framework, we have extended the Lindhard
function of a normal-metal HEG to the superconducting ground state,
thereby constructing the dielectric function of a BCS superconductor.
Our analysis delineates a clear regime of validity for the Lindhard
dielectric function: for energy transfers above $5\Delta$, the normal-state
Lindhard function provides an accurate description of the in-medium
response of superconducting targets.
\begin{figure}[b]
\begin{centering}
\includegraphics[scale=0.27]{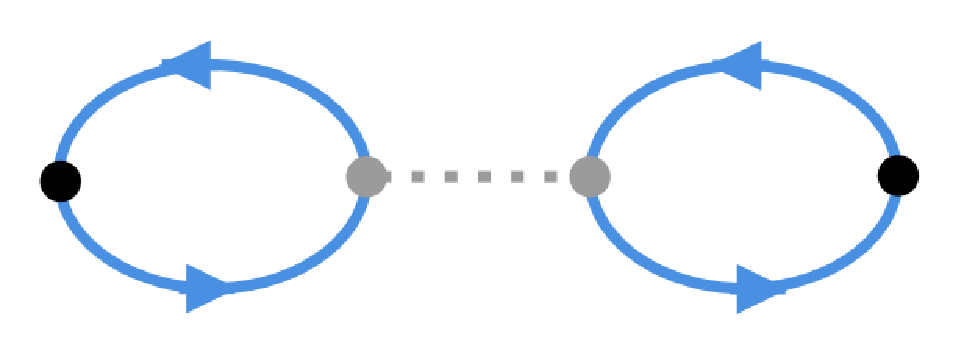}
\par\end{centering}
\caption{An extra polarization function $\Pi_{\theta}$ is introduced by the
phase field $\theta$, where the gray circle represents the $\theta$-$e$
vertex $\propto\Delta\theta\psi^{\dagger}_{e}\tau^{2}\psi_{e}$, and
the gray dotted line denotes the dressed propagator of the phase field
$\theta$.}\label{fig:theta_diagram}

\end{figure}

Yet further extension is possible. Throughout this study we have described
the target at the BCS mean-field level, which is expected to be adequate
for the QP-excitation processes induced by DM scattering at a finite
$Q$. In the long-wavelength limit, however, the collective dynamics
of the condensate phase becomes prominent and calls for a description
beyond mean-field theory. One must then reinstate the phase degree
of freedom by restoring the phase of the order parameter $\Delta\rightarrow\Delta e^{i\theta}$,
which gives rise to a phase field-electron coupling term $\propto\Delta\theta\psi^{\dagger}_{e}\tau^{2}\psi_{e}$,
with $\tau^{2}=\left(\begin{array}{cc}
0 & -i\\
i & 0
\end{array}\right)$. Consequently, in addition to the QP bubble $\Pi_{\mathrm{QP}}$,
the response of the target acquires a further contribution $\Pi_{\theta}$
describing the phase exchange. Dimensional analysis of the corresponding
diagram in Fig~\ref{fig:theta_diagram} suggests that $\Pi_{\theta}$
is parametrically suppressed with respect to $\Pi_{\mathrm{QP}}$
by $\Delta^{2}/\left(v_{F}Q\right)^{2}\sim\left(\xi^{-1}_{0}/Q\right)^{2}$.
The inverse coherence length $\xi^{-1}_{0}=\pi\Delta/v_{F}$ thus
emerges as a natural crossover scale separating the mean-field, QP
regime ($Q\gg1/\xi_{0}$) from the collective, hydrodynamic regime
of the superconducting state ($Q\ll1/\xi_{0}$). While the mean-field
approach is sufficient for the direct detection of halo DM with energy
transfers above $5\Delta$, where the relative error $\Delta^{2}/\left(v_{F}Q\right)^{2}$
is at most $1/25$ (see Fig.~\ref{fig:Ratios}), an accurate description
of many-body physics in superconducting targets at the crossover scale
becomes essential, for the regime approaching the pair-breaking threshold
$\omega\rightarrow2\Delta$. A detailed analysis of this issue is
left for future work.
\begin{acknowledgments}
\textit{Acknowledgements}. Z.L.L. is supported by the National Natural
Science Foundation of China under Grant No.~12575117. F.Z. is supported
by the National Natural Science Foundation of China under Grant No.~12374054.
\end{acknowledgments}

\appendix

\section{ Detailed BCS Formalism and Bogoliubov Quasiparticles}

\label{sec:app-bcs}

This appendix provides a self-contained review of the BCS theoretical
framework used in the main text, presented within the Nambu spinor
formalism.

\subsection{Coulomb Interaction in the Nambu Formalism}

We begin with the Coulomb interaction between electrons, which constitutes
the fundamental electronic interaction in condensed matter systems.
In real space, the Coulomb interaction between position $\mathbf{x}$
and $\mathbf{x}'$ is given by the Lagrangian
\begin{equation}
\begin{split}\int\mbox{d}^{4}x\,\mathcal{L}_{ee} & =-\frac{1}{2}\int\mbox{d}^{4}x\,\mbox{d}^{3}x'\,\psi^{\dagger}_{e}(\mathbf{x},t)\psi_{e}(\mathbf{x},t)\\
 & \quad\times V^{\mathrm{C}}(\mathbf{x}-\mathbf{x}')\psi^{\dagger}_{e}(\mathbf{x}',t)\psi_{e}(\mathbf{x}',t),
\end{split}
\label{eq:A-Coulomb}
\end{equation}
where $V^{\mathrm{C}}(\mathbf{x}-\mathbf{x}')=\frac{e^{2}}{4\pi|\mathbf{x}-\mathbf{x}'|}$
denotes the Coulomb potential. Within the BCS framework, we decompose
the free electron field in the plane-wave basis, \emph{i.e.}, $\psi_{e}(\mathbf{x},t)=\sum_{\mathbf{p},\sigma}c_{\mathbf{p},\sigma}(t)e^{i\mathbf{p}\cdot\mathbf{x}}/\sqrt{V}$,
where $c_{\mathbf{p},\sigma}$ is the annihilation operator for an
electron with momentum $\mathbf{p}$ and spin $\sigma=\uparrow,\downarrow$,
and $V$ denotes the volume of the detector. The electron density
operator then becomes $\psi^{\dagger}_{e}(\mathbf{x},t)\psi_{e}(\mathbf{x},t)=\sum_{\mathbf{p},\mathbf{k},\sigma}c^{\dagger}_{\mathbf{p},\sigma}(t)c_{\mathbf{p}+\mathbf{k},\sigma}(t)e^{i\mathbf{k}\cdot\mathbf{x}}/V$.
Substituting this into Eq.~(\ref{eq:A-Coulomb}) and performing the
spatial integrals yields the momentum-space representation of the
four-fermion interaction:
\begin{equation}
\begin{aligned}[b]\int\mbox{d}^{4}x\,\mathcal{L}_{ee} & =-\frac{1}{2V}\sum_{\mathbf{p},\sigma}\sum_{\mathbf{p}',\sigma'}\sum_{\mathbf{k}}\int\mbox{d}t\ c^{\dagger}_{\mathbf{p},\sigma}c_{\mathbf{p}+\mathbf{k},\sigma}\\
 & \quad\times V^{\mathrm{C}}(-\mathbf{k})c^{\dagger}_{\mathbf{p}',\sigma'}c_{\mathbf{p}'-\mathbf{k},\sigma'},
\end{aligned}
\label{eq:A-Coulomb_k}
\end{equation}
with $V^{\mathrm{C}}(-\mathbf{k})=4\pi\alpha/k^{2}$ ($k=|\mathbf{k}|$),
where $\alpha=e^{2}/(4\pi)$ reprensents the electromagnetic fine-structure
constant.

To treat the electron-hole mixing induced by superconductivity, it
is convenient to adopt Nambu's spinor formulation. The Nambu spinor
is defined as $\psi_{\mathbf{p}}=(c_{\mathbf{p},\uparrow},c^{\dagger}_{-\mathbf{p},\downarrow})^{T}$,
whose components represent an electron and a hole, respectively, and
the corresponding Hermitian conjugate is $\psi^{\dagger}_{\mathbf{p}}=(c^{\dagger}_{\mathbf{p},\uparrow},c_{-\mathbf{p},\downarrow})$.
In this basis, the density operator takes a compact form: $\sum_{\mathbf{p},\sigma}c^{\dagger}_{\mathbf{p},\sigma}c_{\mathbf{p}+\mathbf{k},\sigma}=\sum_{\mathbf{p}}\psi^{\dagger}_{\mathbf{p}-\mathbf{k}}\tau^{3}\psi_{\mathbf{p}}$,
where $\tau^{3}=\operatorname{diag}(1,-1)$ is the third Pauli matrix
in isospin space. The diagonal entries $\pm1$ reflect the opposite
electric charge of electron and hole components. Substituting this
identity into Eq.~(\ref{eq:A-Coulomb_k}), the Coulomb interaction
in Nambu notation reads
\begin{equation}
\begin{aligned}[b]\int\mbox{d}^{4}x\,\mathcal{L}_{ee} & =-\frac{1}{2V}\sum_{\mathbf{p},\mathbf{p}',\mathbf{k}}\int\mbox{d}t\,\psi^{\dagger}_{\mathbf{p}-\mathbf{k}}\tau^{3}\psi_{\mathbf{p}}\\
 & \quad\times\,V^{\mathrm{C}}(-\mathbf{k})\psi^{\dagger}_{\mathbf{p}'+\mathbf{k}}\tau^{3}\psi_{\mathbf{p}'}.
\end{aligned}
\label{eq:A-Coulomb_Nambu}
\end{equation}
Equation~(\ref{eq:A-Coulomb_Nambu}) serves as the starting point
for incorporating superconducting correlations: in the following subsection,
the electron field $\psi_{\mathbf{p}}$ will be transformed to the
quasiparticle basis via the Bogoliubov rotation.

\subsection{Bogoliubov Transformation}

In the BCS mean-field theory, the presence of a superconducting condensate
modifies the electron dynamics through pairing correlations. The electron
Lagrangian, after integrating out the phonon-mediated attractive interaction
and performing the mean-field decoupling, takes the form
\begin{equation}
\int\mbox{d}^{4}x\,\mathcal{L}_{e}=\int\mbox{d}t\sum_{\mathbf{p}}\psi^{\dagger}_{\mathbf{p}}\!\left[i\frac{\partial}{\partial t}-\begin{pmatrix}\varepsilon_{\mathbf{p}} & \Delta\\
\Delta & -\varepsilon_{\mathbf{p}}
\end{pmatrix}\right]\!\psi_{\mathbf{p}},\label{eq:A-BCS_action}
\end{equation}
where $\varepsilon_{\mathbf{p}}=|\mathbf{p}|^{2}/(2m_{e})-E_{F}$
is the single-particle energy measured from the Fermi level, and $\Delta$
is the superconducting gap energy. The off-diagonal elements $\Delta$
couple the electron component ($c_{\mathbf{p},\uparrow}$) to the
hole component ($c^{\dagger}_{-\mathbf{p},\downarrow}$), reflecting
the fact that the BCS ground state is a coherent superposition of
electron pairs $(\mathbf{p}\uparrow,-\mathbf{p}\downarrow)$. The
gap $\Delta$ is not a free parameter but is determined self-consistently
from the BCS gap equation, which encodes the balance between the attractive
phonon-mediated interaction and the repulsive Coulomb pseudopotential.

Because the $2\times2$ Hamiltonian matrix in Eq.~(\ref{eq:A-BCS_action})
is non-diagonal, the electron field $\psi_{\mathbf{p}}$ is no longer
an energy eigenstate of the superconducting system. The proper degrees
of freedom are the Bogoliubov \emph{quasiparticles} (QPs), obtained
by diagonalizing the Hamiltonian matrix. We introduce the Bogoliubov
unitary matrix $U_{\mathbf{p}}$ satisfying
\begin{equation}
U^{\dagger}_{\mathbf{p}}\begin{pmatrix}\varepsilon_{\mathbf{p}} & \Delta\\
\Delta & -\varepsilon_{\mathbf{p}}
\end{pmatrix}U_{\mathbf{p}}=\begin{pmatrix}E_{\mathbf{p}} & 0\\
0 & -E_{\mathbf{p}}
\end{pmatrix},
\end{equation}
where $E_{\mathbf{p}}=\sqrt{\varepsilon^{2}_{\mathbf{p}}+\Delta^{2}}$
is the QP energy dispersion. The eigenvalue problem yields the parameterization
$U_{\mathbf{p}}=\left(\begin{array}{cc}
u_{\mathbf{p}} & -v_{\mathbf{p}}\\
v_{\mathbf{p}} & u_{\mathbf{p}}
\end{array}\right)$, with the coherence factors
\begin{equation}
u_{\mathbf{p}}=\sqrt{\frac{1}{2}\!\left(1+\frac{\varepsilon_{\mathbf{p}}}{E_{\mathbf{p}}}\right)},\qquad v_{\mathbf{p}}=\sqrt{\frac{1}{2}\!\left(1-\frac{\varepsilon_{\mathbf{p}}}{E_{\mathbf{p}}}\right)}.
\end{equation}
Physically, $v^{2}_{\mathbf{p}}$ gives the probability that the pair
state $(\mathbf{p}\uparrow,-\mathbf{p}\downarrow)$ is occupied in
the BCS ground state, while $u^{2}_{\mathbf{p}}=1-v^{2}_{\mathbf{p}}$
is the probability that it is empty. Far above the Fermi surface ($\varepsilon_{\mathbf{p}}\gg\Delta$),
one has $u_{\mathbf{p}}\approx1$, $v_{\mathbf{p}}\approx0$, and
the QPs reduce to ordinary electrons; far below ($-\varepsilon_{\mathbf{p}}\gg\Delta$),
$u_{\mathbf{p}}\approx0$, $v_{\mathbf{p}}\approx1$, and they become
holes.

The QP field $\Psi_{\mathbf{p}}=U^{\dagger}_{\mathbf{p}}\psi_{\mathbf{p}}=(\gamma_{\mathbf{p},\uparrow},\gamma^{\dagger}_{-\mathbf{p},\downarrow})^{T}$
diagonalizes the BCS action:
\begin{equation}
\int\mbox{d}^{4}x\,\mathcal{L}_{e}=\int\mbox{d}t\sum_{\mathbf{p}}\Psi^{\dagger}_{\mathbf{p}}\!\left[i\frac{\partial}{\partial t}-\begin{pmatrix}E_{\mathbf{p}} & 0\\
0 & -E_{\mathbf{p}}
\end{pmatrix}\right]\!\Psi_{\mathbf{p}}.\label{eq:A-QP_action}
\end{equation}
The inverse Bogoliubov transformation relates the QP operators to
the original electron operators,
\begin{equation}
\begin{pmatrix}\gamma_{\mathbf{p},\uparrow}\\
\gamma^{\dagger}_{-\mathbf{p},\downarrow}
\end{pmatrix}=\begin{pmatrix}u_{\mathbf{p}} & v_{\mathbf{p}}\\
-v_{\mathbf{p}} & u_{\mathbf{p}}
\end{pmatrix}\begin{pmatrix}c_{\mathbf{p},\uparrow}\\
c^{\dagger}_{-\mathbf{p},\downarrow}
\end{pmatrix},\label{eq:A-QP_transform}
\end{equation}
which explicitly reveals the QP as a coherent superposition of an
electron and a hole: the annihilation of a QP ($\gamma_{\mathbf{p},\uparrow}$)
corresponds to removing an electron with amplitude $u_{\mathbf{p}}$
or adding a hole with amplitude $v_{\mathbf{p}}$, reflecting the
inherent particle-hole mixing in the superconducting state.
\begin{widetext}
From Eq.~(\ref{eq:A-QP_action}), one can obtain the corresponding
QP propagator
\begin{eqnarray}
S_{\mathbf{p}}\left(t-t'\right) & = & \left(\begin{array}{cc}
\left\langle \hat{\gamma}_{\mathbf{p},\uparrow}\left(t\right)\hat{\gamma}^{\dagger}_{\mathbf{p},\uparrow}\left(t'\right)\right\rangle  & \left\langle \hat{\gamma}_{\mathbf{p},\uparrow}\left(t\right)\hat{\gamma}_{-\mathbf{p},\downarrow}\left(t'\right)\right\rangle \\
\left\langle \hat{\gamma}^{\dagger}_{-\mathbf{p},\downarrow}\left(t\right)\hat{\gamma}^{\dagger}_{\mathbf{p},\uparrow}\left(t'\right)\right\rangle  & \left\langle \hat{\gamma}^{\dagger}_{-\mathbf{p},\downarrow}\left(t\right)\hat{\gamma}_{-\mathbf{p},\downarrow}\left(t'\right)\right\rangle
\end{array}\right)\nonumber \\
 & = & \int S_{\mathbf{p}}\left(\omega\right)\frac{e^{-i\omega\left(t-t'\right)}\mathrm{d}\omega}{2\pi}\nonumber \\
 & = & \int\left(\begin{array}{cc}
\frac{1}{\omega-E_{\mathbf{p}}+i0^{+}} & 0\\
0 & \frac{1}{\omega+E_{\mathbf{p}}-i0^{+}}
\end{array}\right)\frac{e^{-i\omega\left(t-t'\right)}\mathrm{d}\omega}{2\pi},\label{eq:QP_progagator}
\end{eqnarray}
where $\langle\cdots\rangle$ denotes the two-time correlation function
in the BCS vacuum at zero temperature, and the infinitesimal $i0^{+}$
stems from the QP one-particle spectral function.

\subsection{DM-Quasiparticle Interaction Vertex}

Having established the QP basis, we now specialize to the DM-electron
interaction. When both the DM coupling and the Coulomb interaction
between electrons are expressed in terms of the QP field $\Psi_{\mathbf{p}}$,
the relevant vertex factor becomes $U^{\dagger}_{\mathbf{p}'}\tau^{3}U_{\mathbf{p}}$,
where the isospin matrix $\tau^{3}$ originates from the electron
density operator in the Nambu formalism. Multiplying this vertex factor
by the QP field operators yields the four-fermion interaction in the
QP basis. Explicit evaluation gives
\begin{align}
\Psi^{\dagger}_{\mathbf{p}'}U^{\dagger}_{\mathbf{p}'}\tau^{3}U_{\mathbf{p}}\Psi_{\mathbf{p}} & =\begin{pmatrix}\gamma^{\dagger}_{\mathbf{p}',\uparrow} & \gamma_{-\mathbf{p}',\downarrow}\end{pmatrix}\begin{pmatrix}u_{\mathbf{p}'}u_{\mathbf{p}}-v_{\mathbf{p}'}v_{\mathbf{p}} & -u_{\mathbf{p}'}v_{\mathbf{p}}-v_{\mathbf{p}'}u_{\mathbf{p}}\\
-u_{\mathbf{p}'}v_{\mathbf{p}}-v_{\mathbf{p}'}u_{\mathbf{p}} & -u_{\mathbf{p}'}u_{\mathbf{p}}+v_{\mathbf{p}'}v_{\mathbf{p}}
\end{pmatrix}\begin{pmatrix}\gamma_{\mathbf{p},\uparrow}\\
\gamma^{\dagger}_{-\mathbf{p},\downarrow}
\end{pmatrix}.\label{eq:A-vertex}
\end{align}
The four matrix elements above correspond, from left to right and
top to bottom, to the vertices for: $\gamma^{\dagger}\gamma$ (QP-preserving),
$\gamma^{\dagger}\gamma^{\dagger}$ (pair creation from vacuum), $\gamma\gamma$
(pair annihilation into vacuum), and $\gamma\gamma^{\dagger}$ (QP-preserving,
hole sector). This structure is the QP-basis analog of the familiar
electron-density vertex in normal metals.

For the two-QP final state $|\mathbf{p}_{1},\uparrow;\mathbf{p}_{2},\downarrow\rangle_{\mathrm{QP}}$
produced when a DM particle breaks a Cooper pair, the LSZ reduction
formula selects only terms in which each outgoing QP is associated
with a creation operator. From Eq.~(\ref{eq:QP_progagator}), only
the pairing pattern $\langle\hat{\gamma}\hat{\gamma}^{\dagger}\rangle$
or $\langle\hat{\gamma}^{\dagger}\hat{\gamma}\rangle$ for the QP
external legs survives in the time-ordered correlation function. Consequently,
only the off-diagonal element $\propto\gamma^{\dagger}_{\mathbf{p}',\uparrow}\gamma^{\dagger}_{-\mathbf{p}',\downarrow}$
(i.e., the term with creation operators for both QP branches) contributes
to the scattering amplitude. The corresponding coefficient is $-(u_{\mathbf{p}_{1}}v_{\mathbf{p}_{2}}+v_{\mathbf{p}_{1}}u_{\mathbf{p}_{2}})$,
which is precisely the BCS coherence factor appearing in the tree-level
$\mathcal{T}$-matrix in the main text. This factor encodes the projection
of the DM-electron density coupling from the electron basis onto the
physical QP eigenstates.

\section{ Lindhard Dielectric Function}

\label{sec:app-lindhard}

For completeness, we present the full Lindhard dielectric function
for a homogeneous electron gas (HEG) at zero temperature~\citep{Lindhard:1954},
which serves as the normal-state reference against which the BCS dielectric
function is compared in the main text. The Lindhard function provides
a closed-form analytic expression for the RPA dielectric response
of a degenerate electron gas, and it has been widely employed in the
analysis of DM direct detection experiments using both normal-metal
and superconducting targets.

In the zero-temperature limit, the Lindhard dielectric function can
be decomposed into real and imaginary parts as follows: {\allowdisplaybreaks
\begin{align}
\epsilon_{\mathrm{Lin}}(\mathbf{Q},\omega) & =1+\frac{4\pi\alpha}{Q^{2}}\frac{2m_{e}p_{F}}{(2\pi)^{2}}\Bigg\{1+\frac{p_{F}}{2Q}\!\left[1-\left(\frac{Q}{2p_{F}}-\frac{\omega}{v_{F}Q}\right)^{2}\right]\log\left|\frac{1+(\frac{Q}{2p_{F}}-\frac{\omega}{v_{F}Q})}{1-(\frac{Q}{2p_{F}}-\frac{\omega}{v_{F}Q})}\right|\nonumber \\
 & \quad+\frac{p_{F}}{2Q}\!\left[1-\left(\frac{Q}{2p_{F}}+\frac{\omega}{v_{F}Q}\right)^{2}\right]\log\left|\frac{1+(\frac{Q}{2p_{F}}+\frac{\omega}{v_{F}Q})}{1-(\frac{Q}{2p_{F}}+\frac{\omega}{v_{F}Q})}\right|\Bigg\}\nonumber \\
 & \quad+i\frac{4\pi\alpha}{Q^{2}}\frac{2m_{e}p_{F}}{(2\pi)^{2}}\times\begin{cases}
\frac{\pi\omega}{v_{F}Q}, & (p_{F}-\frac{Q}{2})\frac{Q}{m_{e}}>\omega\\[4pt]
\frac{\pi p_{F}}{2Q}\!\left[1-\left(\frac{Q}{2p_{F}}-\frac{\omega}{v_{F}Q}\right)^{2}\right], & (p_{F}+\frac{Q}{2})\frac{Q}{m_{e}}>\omega>(-p_{F}+\frac{Q}{2})\frac{Q}{m_{e}}\\[4pt]
0, & \omega>(p_{F}+\frac{Q}{2})\frac{Q}{m_{e}}\\
0, & \omega<(-p_{F}+\frac{Q}{2})\frac{Q}{m_{e}}
\end{cases}\label{eq:A-Lindhard}
\end{align}
}
\end{widetext}

The imaginary part encodes the dissipative response associated with
electron-hole pair excitations, and is non-zero only within the kinematically
allowed particle-hole continuum bounded by the curves $\omega=(Q/2\pm p_{F})Q/m_{e}$.
These two boundary curves, plotted as red dotted lines in Fig.~2
of the main text, mark the region where the Lindhard function can
absorb energy from an external probe via single-particle excitations.

In Eq.~(\ref{eq:A-Lindhard}), $p_{F}$ and $v_{F}=p_{F}/m_{e}$
denote the Fermi momentum and Fermi velocity, respectively. For the
materials considered in this work, the relevant parameters are $p_{F}=3.44\,\mathrm{keV}$
and $v_{F}=6.75\times10^{-3}c$ for aluminum, and $p_{F}=2.67\,\mathrm{keV}$
and $v_{F}=5.24\times10^{-3}c$ for WSi.

\bibliographystyle{JHEP1}
\addcontentsline{toc}{section}{\refname}\bibliography{dielectric_function_superconductor_arXiv.bbl}

\providecommand{\href}[2]{#2}\begingroup\raggedright\begin{thebibliography}{10}

\bibitem{Essig:2011nj}
R.~Essig, J.~Mardon and T.~Volansky, \emph{{Direct Detection of Sub-GeV Dark
  Matter}}, \href{https://doi.org/10.1103/PhysRevD.85.076007}{\emph{Phys. Rev.}
  {\bfseries D85} (2012) 076007}
  [\href{https://arxiv.org/abs/1108.5383}{{\ttfamily 1108.5383}}].

\bibitem{Graham:2012su}
P.~W. Graham, D.~E. Kaplan, S.~Rajendran and M.~T. Walters,
  \emph{{Semiconductor Probes of Light Dark Matter}},
  \href{https://doi.org/10.1016/j.dark.2012.09.001}{\emph{Phys. Dark Univ.}
  {\bfseries 1} (2012) 32} [\href{https://arxiv.org/abs/1203.2531}{{\ttfamily
  1203.2531}}].

\bibitem{Hochberg:2015fth}
Y.~Hochberg, M.~Pyle, Y.~Zhao and K.~M. Zurek, \emph{{Detecting Superlight Dark
  Matter with Fermi-Degenerate Materials}},
  \href{https://doi.org/10.1007/JHEP08(2016)057}{\emph{JHEP} {\bfseries 08}
  (2016) 057} [\href{https://arxiv.org/abs/1512.04533}{{\ttfamily
  1512.04533}}].

\bibitem{Essig:2015cda}
R.~Essig, M.~Fernandez-Serra, J.~Mardon, A.~Soto, T.~Volansky and T.-T. Yu,
  \emph{{Direct Detection of sub-GeV Dark Matter with Semiconductor Targets}},
  \href{https://doi.org/10.1007/JHEP05(2016)046}{\emph{JHEP} {\bfseries 05}
  (2016) 046} [\href{https://arxiv.org/abs/1509.01598}{{\ttfamily
  1509.01598}}].

\bibitem{Griffin:2018bjn}
S.~Griffin, S.~Knapen, T.~Lin and K.~M. Zurek, \emph{{Directional Detection of
  Light Dark Matter with Polar Materials}},
  \href{https://doi.org/10.1103/PhysRevD.98.115034}{\emph{Phys. Rev.}
  {\bfseries D98} (2018) 115034}
  [\href{https://arxiv.org/abs/1807.10291}{{\ttfamily 1807.10291}}].

\bibitem{Ibe2018}
M.~Ibe, W.~Nakano, Y.~Shoji and K.~Suzuki, \emph{Migdal effect in dark matter
  direct detection experiments},
  \href{https://doi.org/10.1007/JHEP03(2018)194}{\emph{Journal of High Energy
  Physics} {\bfseries 2018} (2018) 194}.

\bibitem{Knapen2018}
S.~Knapen, T.~Lin, M.~Pyle and K.~M. Zurek, \emph{{Detection of Light Dark
  Matter With Optical Phonons in Polar Materials}},
  \href{https://doi.org/10.1016/j.physletb.2018.08.064}{\emph{Phys. Lett.}
  {\bfseries B785} (2018) 386}
  [\href{https://arxiv.org/abs/1712.06598}{{\ttfamily 1712.06598}}].

\bibitem{Liang:2018bdb}
Z.-L. Liang, L.~Zhang, P.~Zhang and F.~Zheng, \emph{{The wavefunction
  reconstruction effects in calculation of DM-induced electronic transition in
  semiconductor targets}},
  \href{https://doi.org/10.1007/JHEP01(2019)149}{\emph{JHEP} {\bfseries 01}
  (2019) 149} [\href{https://arxiv.org/abs/1810.13394}{{\ttfamily
  1810.13394}}].

\bibitem{Campbell-Deem:2019hdx}
B.~Campbell-Deem, P.~Cox, S.~Knapen, T.~Lin and T.~Melia, \emph{{Multiphonon
  excitations from dark matter scattering in crystals}},
  \href{https://doi.org/10.1103/PhysRevD.101.036006}{\emph{Phys. Rev. D}
  {\bfseries 101} (2020) 036006}
  [\href{https://arxiv.org/abs/1911.03482}{{\ttfamily 1911.03482}}].

\bibitem{Geilhufe:2019ndy}
R.~M. Geilhufe, F.~Kahlhoefer and M.~W. Winkler, \emph{{Dirac Materials for
  Sub-MeV Dark Matter Detection: New Targets and Improved Formalism}},
  \href{https://doi.org/10.1103/PhysRevD.101.055005}{\emph{Phys. Rev. D}
  {\bfseries 101} (2020) 055005}
  [\href{https://arxiv.org/abs/1910.02091}{{\ttfamily 1910.02091}}].

\bibitem{Griffin:2019mvc}
S.~M. Griffin, K.~Inzani, T.~Trickle, Z.~Zhang and K.~M. Zurek,
  \emph{{Multichannel direct detection of light dark matter: Target
  comparison}}, \href{https://doi.org/10.1103/PhysRevD.101.055004}{\emph{Phys.
  Rev. D} {\bfseries 101} (2020) 055004}
  [\href{https://arxiv.org/abs/1910.10716}{{\ttfamily 1910.10716}}].

\bibitem{Liang:2019nnx}
Z.-L. Liang, L.~Zhang, F.~Zheng and P.~Zhang, \emph{{Describing Migdal effects
  in diamond crystal with atom-centered localized Wannier functions}},
  \href{https://doi.org/10.1103/PhysRevD.102.043007}{\emph{Phys. Rev. D}
  {\bfseries 102} (2020) 043007}
  [\href{https://arxiv.org/abs/1912.13484}{{\ttfamily 1912.13484}}].

\bibitem{Trickle:2019nya}
T.~Trickle, Z.~Zhang, K.~M. Zurek, K.~Inzani and S.~Griffin,
  \emph{{Multi-Channel Direct Detection of Light Dark Matter: Theoretical
  Framework}}, \href{https://doi.org/10.1007/JHEP03(2020)036}{\emph{JHEP}
  {\bfseries 03} (2020) 036}
  [\href{https://arxiv.org/abs/1910.08092}{{\ttfamily 1910.08092}}].

\bibitem{Kahn:2020fef}
Y.~Kahn, G.~Krnjaic and B.~Mandava, \emph{{Dark Matter Detection with Bound
  Nuclear Targets: The Poisson Phonon Tail}},
  \href{https://doi.org/10.1103/PhysRevLett.127.081804}{\emph{Phys. Rev. Lett.}
  {\bfseries 127} (2021) 081804}
  [\href{https://arxiv.org/abs/2011.09477}{{\ttfamily 2011.09477}}].

\bibitem{Knapen:2020aky}
S.~Knapen, J.~Kozaczuk and T.~Lin, \emph{{Migdal Effect in Semiconductors}},
  \href{https://doi.org/10.1103/PhysRevLett.127.081805}{\emph{Phys. Rev. Lett.}
  {\bfseries 127} (2021) 081805}
  [\href{https://arxiv.org/abs/2011.09496}{{\ttfamily 2011.09496}}].

\bibitem{Kozaczuk:2020uzb}
J.~Kozaczuk and T.~Lin, \emph{{Plasmon production from dark matter
  scattering}}, \href{https://doi.org/10.1103/PhysRevD.101.123012}{\emph{Phys.
  Rev. D} {\bfseries 101} (2020) 123012}
  [\href{https://arxiv.org/abs/2003.12077}{{\ttfamily 2003.12077}}].

\bibitem{Kurinsky:2020dpb}
N.~Kurinsky, D.~Baxter, Y.~Kahn and G.~Krnjaic, \emph{{Dark matter
  interpretation of excesses in multiple direct detection experiments}},
  \href{https://doi.org/10.1103/PhysRevD.102.015017}{\emph{Phys. Rev. D}
  {\bfseries 102} (2020) 015017}
  [\href{https://arxiv.org/abs/2002.06937}{{\ttfamily 2002.06937}}].

\bibitem{Trickle:2020oki}
T.~Trickle, Z.~Zhang and K.~M. Zurek, \emph{{Effective field theory of dark
  matter direct detection with collective excitations}},
  \href{https://doi.org/10.1103/PhysRevD.105.015001}{\emph{Phys. Rev. D}
  {\bfseries 105} (2022) 015001}
  [\href{https://arxiv.org/abs/2009.13534}{{\ttfamily 2009.13534}}].

\bibitem{Griffin:2021znd}
S.~M. Griffin, K.~Inzani, T.~Trickle, Z.~Zhang and K.~M. Zurek, \emph{{Extended
  calculation of dark matter-electron scattering in crystal targets}},
  \href{https://doi.org/10.1103/PhysRevD.104.095015}{\emph{Phys. Rev. D}
  {\bfseries 104} (2021) 095015}
  [\href{https://arxiv.org/abs/2105.05253}{{\ttfamily 2105.05253}}].

\bibitem{Vahsen:2021gnb}
S.~E. Vahsen, C.~A.~J. O'Hare and D.~Loomba, \emph{{Directional Recoil
  Detection}},
  \href{https://doi.org/10.1146/annurev-nucl-020821-035016}{\emph{Ann. Rev.
  Nucl. Part. Sci.} {\bfseries 71} (2021) 189}
  [\href{https://arxiv.org/abs/2102.04596}{{\ttfamily 2102.04596}}].

\bibitem{Catena:2021qsr}
R.~Catena, T.~Emken, M.~Matas, N.~A. Spaldin and E.~Urdshals, \emph{{Crystal
  responses to general dark matter-electron interactions}},
  \href{https://arxiv.org/abs/2105.02233}{{\ttfamily 2105.02233}}.

\bibitem{Hochberg:2021pkt}
Y.~Hochberg, Y.~Kahn, N.~Kurinsky, B.~V. Lehmann, T.~C. Yu and K.~K. Berggren,
  \emph{{Determining Dark-Matter\textendash{}Electron Scattering Rates from the
  Dielectric Function}},
  \href{https://doi.org/10.1103/PhysRevLett.127.151802}{\emph{Phys. Rev. Lett.}
  {\bfseries 127} (2021) 151802}
  [\href{https://arxiv.org/abs/2101.08263}{{\ttfamily 2101.08263}}].

\bibitem{Knapen:2021run}
S.~Knapen, J.~Kozaczuk and T.~Lin, \emph{{Dark matter-electron scattering in
  dielectrics}}, \href{https://doi.org/10.1103/PhysRevD.104.015031}{\emph{Phys.
  Rev. D} {\bfseries 104} (2021) 015031}
  [\href{https://arxiv.org/abs/2101.08275}{{\ttfamily 2101.08275}}].

\bibitem{Wang:2021oha}
W.~Wang, K.-Y. Wu, L.~Wu and B.~Zhu, \emph{{Direct Detection of Spin-Dependent
  Sub-GeV Dark Matter via Migdal Effect}},
  \href{https://arxiv.org/abs/2112.06492}{{\ttfamily 2112.06492}}.

\bibitem{Knapen:2021bwg}
S.~Knapen, J.~Kozaczuk and T.~Lin, \emph{{python package for dark matter
  scattering in dielectric targets}},
  \href{https://doi.org/10.1103/PhysRevD.105.015014}{\emph{Phys. Rev. D}
  {\bfseries 105} (2022) 015014}
  [\href{https://arxiv.org/abs/2104.12786}{{\ttfamily 2104.12786}}].

\bibitem{Liang:2022xbu}
Z.-L. Liang, C.~Mo, F.~Zheng and P.~Zhang, \emph{{Phonon-mediated Migdal effect
  in semiconductor detectors}},
  \href{https://doi.org/10.1103/PhysRevD.106.043004}{\emph{Phys. Rev. D}
  {\bfseries 106} (2022) 043004}
  [\href{https://arxiv.org/abs/2205.03395}{{\ttfamily 2205.03395}}].

\bibitem{Berghaus:2022pbu}
K.~V. Berghaus, A.~Esposito, R.~Essig and M.~Sholapurkar, \emph{{The Migdal
  effect in semiconductors for dark matter with masses below \ensuremath{\sim}
  100 MeV}}, \href{https://doi.org/10.1007/JHEP01(2023)023}{\emph{JHEP}
  {\bfseries 01} (2023) 023}
  [\href{https://arxiv.org/abs/2210.06490}{{\ttfamily 2210.06490}}].

\bibitem{Campbell-Deem:2022fqm}
B.~Campbell-Deem, S.~Knapen, T.~Lin and E.~Villarama, \emph{{Dark matter direct
  detection from the single phonon to the nuclear recoil regime}},
  \href{https://arxiv.org/abs/2205.02250}{{\ttfamily 2205.02250}}.

\bibitem{Li:2022acp}
J.~Li, L.~Su, L.~Wu and B.~Zhu, \emph{{Spin-dependent sub-GeV inelastic dark
  matter-electron scattering and Migdal effect. Part I. Velocity independent
  operator}}, \href{https://doi.org/10.1088/1475-7516/2023/04/020}{\emph{JCAP}
  {\bfseries 04} (2023) 020}
  [\href{https://arxiv.org/abs/2210.15474}{{\ttfamily 2210.15474}}].

\bibitem{Esposito:2022bnu}
A.~Esposito and S.~Pavaskar, \emph{{Optimal anti-ferromagnets for light dark
  matter detection}},  \href{https://arxiv.org/abs/2210.13516}{{\ttfamily
  2210.13516}}.

\bibitem{Dreyer:2023ovn}
C.~E. Dreyer, R.~Essig, M.~Fernandez-Serra, A.~Singal and C.~Zhen, \emph{{Fully
  ab-initio all-electron calculation of dark matter--electron scattering in
  crystals with evaluation of systematic uncertainties}},
  \href{https://arxiv.org/abs/2306.14944}{{\ttfamily 2306.14944}}.

\bibitem{Liang:2024xcx}
Z.-L. Liang, L.~Su, L.~Wu and B.~Zhu, \emph{{Plasmon-enhanced Direct Detection
  of sub-MeV Dark Matter}},
  \href{https://doi.org/10.1103/PhysRevLett.134.071001}{\emph{Phys. Rev. Lett.}
  {\bfseries 134} (2025) 071001}
  [\href{https://arxiv.org/abs/2401.11971}{{\ttfamily 2401.11971}}].

\bibitem{Essig:2024ebk}
R.~Essig, R.~Plestid and A.~Singal, \emph{{Collective excitations and
  low-energy ionization signatures of relativistic particles in silicon
  detectors}}, \href{https://doi.org/10.1038/s42005-024-01904-2}{\emph{Commun.
  Phys.} {\bfseries 7} (2024) 416}
  [\href{https://arxiv.org/abs/2403.00123}{{\ttfamily 2403.00123}}].

\bibitem{Dent:2025drd}
J.~B. Dent, B.~A. Friedman, J.~L. Newstead and S.~Sabharwal, \emph{{Nuclear and
  electron scattering by neutrinos and dark matter in condensed systems}},
  \href{https://doi.org/10.1103/plgb-cbk5}{\emph{Phys. Rev. D} {\bfseries 113}
  (2026) 116016} [\href{https://arxiv.org/abs/2510.06574}{{\ttfamily
  2510.06574}}].

\bibitem{Liang:2025nan}
Z.-L. Liang and F.~Zheng, \emph{{Cutting rules for non-relativistic dark matter
  in solids based on Kohn-Sham orbitals}},
  \href{https://doi.org/10.1007/JHEP01(2026)101}{\emph{JHEP} {\bfseries 01}
  (2026) 101} [\href{https://arxiv.org/abs/2507.11033}{{\ttfamily
  2507.11033}}].

\bibitem{Esposito:2025iry}
A.~Esposito and A.~Rocchi, \emph{{Migdal effect in solid crystals and the role
  of nonadiabaticity}}, \href{https://doi.org/10.1103/198q-nv94}{\emph{Phys.
  Rev. D} {\bfseries 112} (2025) 075034}
  [\href{https://arxiv.org/abs/2505.08864}{{\ttfamily 2505.08864}}].

\bibitem{Mai:2025zau}
Q.~Mai, G.~Huai-Ke and Z.~Yu-Feng, \emph{{Progresses of theory on underground
  light dark matter direct detection{\textemdash}Cosmic-ray boosted dark matter
  and the Migdal effect}},
  \href{https://doi.org/10.1360/SSPMA-2024-0485}{\emph{Sci. Sin. Phys. Mech.
  Astro.} {\bfseries 55} (2025) 111006}.

\bibitem{Sun:2025gyj}
J.-W. Sun, L.~Wu, Y.-H. Xu and B.~Zhu, \emph{{Probing supernova neutrino
  boosted dark matter with collective excitations}},
  \href{https://doi.org/10.1103/xhl3-v17n}{\emph{Phys. Rev. D} {\bfseries 112}
  (2025) 015014} [\href{https://arxiv.org/abs/2501.07591}{{\ttfamily
  2501.07591}}].

\bibitem{Amaral:2020ryn}
{\scshape SuperCDMS} collaboration, \emph{{Constraints on low-mass, relic dark
  matter candidates from a surface-operated SuperCDMS single-charge sensitive
  detector}}, \href{https://doi.org/10.1103/PhysRevD.102.091101}{\emph{Phys.
  Rev. D} {\bfseries 102} (2020) 091101}
  [\href{https://arxiv.org/abs/2005.14067}{{\ttfamily 2005.14067}}].

\bibitem{CRESST:2019jnq}
{\scshape CRESST} collaboration, \emph{{First results from the CRESST-III
  low-mass dark matter program}},
  \href{https://doi.org/10.1103/PhysRevD.100.102002}{\emph{Phys. Rev. D}
  {\bfseries 100} (2019) 102002}
  [\href{https://arxiv.org/abs/1904.00498}{{\ttfamily 1904.00498}}].

\bibitem{Arnaud:2020svb}
{\scshape EDELWEISS} collaboration, \emph{{First germanium-based constraints on
  sub-MeV Dark Matter with the EDELWEISS experiment}},
  \href{https://doi.org/10.1103/PhysRevLett.125.141301}{\emph{Phys. Rev. Lett.}
  {\bfseries 125} (2020) 141301}
  [\href{https://arxiv.org/abs/2003.01046}{{\ttfamily 2003.01046}}].

\bibitem{Barak:2020fql}
{\scshape SENSEI} collaboration, \emph{{SENSEI: Direct-Detection Results on
  sub-GeV Dark Matter from a New Skipper-CCD}},
  \href{https://doi.org/10.1103/PhysRevLett.125.171802}{\emph{Phys. Rev. Lett.}
  {\bfseries 125} (2020) 171802}
  [\href{https://arxiv.org/abs/2004.11378}{{\ttfamily 2004.11378}}].

\bibitem{Castello-Mor:2020jhd}
{\scshape DAMIC-M} collaboration, \emph{{DAMIC-M Experiment: Thick, Silicon
  CCDs to search for Light Dark Matter}},
  \href{https://doi.org/10.1016/j.nima.2019.162933}{\emph{Nucl. Instrum. Meth.
  A} {\bfseries 958} (2020) 162933}
  [\href{https://arxiv.org/abs/2001.01476}{{\ttfamily 2001.01476}}].

\bibitem{COSINE-100:2021poy}
{\scshape COSINE-100} collaboration, \emph{{Searching for low-mass dark matter
  via the Migdal effect in COSINE-100}},
  \href{https://doi.org/10.1103/PhysRevD.105.042006}{\emph{Phys. Rev. D}
  {\bfseries 105} (2022) 042006}
  [\href{https://arxiv.org/abs/2110.05806}{{\ttfamily 2110.05806}}].

\bibitem{SuperCDMS:2022kgp}
{\scshape SuperCDMS} collaboration, \emph{{A Search for Low-mass Dark Matter
  via Bremsstrahlung Radiation and the Migdal Effect in SuperCDMS}},
  \href{https://arxiv.org/abs/2203.02594}{{\ttfamily 2203.02594}}.

\bibitem{CDEX:2022kcd}
{\scshape CDEX} collaboration, \emph{{Constraints on Sub-GeV Dark
  Matter\textendash{}Electron Scattering from the CDEX-10 Experiment}},
  \href{https://doi.org/10.1103/PhysRevLett.129.221301}{\emph{Phys. Rev. Lett.}
  {\bfseries 129} (2022) 221301}
  [\href{https://arxiv.org/abs/2206.04128}{{\ttfamily 2206.04128}}].

\bibitem{CONNIE:2024off}
{\scshape CONNIE, Atucha-II} collaboration, \emph{{Search for Reactor-Produced
  Millicharged Particles with Skipper-CCDs at the CONNIE and Atucha-II
  Experiments}},
  \href{https://doi.org/10.1103/PhysRevLett.134.071801}{\emph{Phys. Rev. Lett.}
  {\bfseries 134} (2025) 071801}
  [\href{https://arxiv.org/abs/2405.16316}{{\ttfamily 2405.16316}}].

\bibitem{Kahn:2021ttr}
Y.~Kahn and T.~Lin, \emph{{Searches for light dark matter using condensed
  matter systems}}, \href{https://doi.org/10.1088/1361-6633/ac5f63}{\emph{Rept.
  Prog. Phys.} {\bfseries 85} (2022) 066901}
  [\href{https://arxiv.org/abs/2108.03239}{{\ttfamily 2108.03239}}].

\bibitem{Hochberg:2015pha}
Y.~Hochberg, Y.~Zhao and K.~M. Zurek, \emph{{Superconducting Detectors for
  Superlight Dark Matter}},
  \href{https://doi.org/10.1103/PhysRevLett.116.011301}{\emph{Phys. Rev. Lett.}
  {\bfseries 116} (2016) 011301}
  [\href{https://arxiv.org/abs/1504.07237}{{\ttfamily 1504.07237}}].

\bibitem{Hochberg:2021ymx}
Y.~Hochberg, E.~D. Kramer, N.~Kurinsky and B.~V. Lehmann, \emph{{Directional
  detection of light dark matter in superconductors}},
  \href{https://doi.org/10.1103/PhysRevD.107.076015}{\emph{Phys. Rev. D}
  {\bfseries 107} (2023) 076015}
  [\href{https://arxiv.org/abs/2109.04473}{{\ttfamily 2109.04473}}].

\bibitem{Hochberg:2021yud}
Y.~Hochberg, B.~V. Lehmann, I.~Charaev, J.~Chiles, M.~Colangelo, S.~W. Nam
  et~al., \emph{{New constraints on dark matter from superconducting
  nanowires}}, \href{https://doi.org/10.1103/PhysRevD.106.112005}{\emph{Phys.
  Rev. D} {\bfseries 106} (2022) 112005}
  [\href{https://arxiv.org/abs/2110.01586}{{\ttfamily 2110.01586}}].

\bibitem{Gao:2024irf}
J.~Gao, Y.~Hochberg, B.~V. Lehmann, S.~W. Nam, P.~Szypryt, M.~R. Vissers
  et~al., \emph{{Detecting Light Dark Matter with Kinetic Inductance
  Detectors}},  \href{https://arxiv.org/abs/2403.19739}{{\ttfamily
  2403.19739}}.

\bibitem{Griffin:2024jec}
S.~M. Griffin, G.~D. Hadas, Y.~Hochberg, K.~Inzani and B.~V. Lehmann,
  \emph{{Dark-Matter{\textendash}Electron Detectors for
  Dark-Matter{\textendash}Nucleon Interactions}},
  \href{https://doi.org/10.1103/6qqv-rl7q}{\emph{Phys. Rev. Lett.} {\bfseries
  135} (2025) 141803} [\href{https://arxiv.org/abs/2412.16283}{{\ttfamily
  2412.16283}}].

\bibitem{Das:2024jdz}
A.~Das, N.~Kurinsky and R.~K. Leane, \emph{{Transmon Qubit constraints on dark
  matter-nucleon scattering}},
  \href{https://doi.org/10.1007/JHEP07(2024)233}{\emph{JHEP} {\bfseries 07}
  (2024) 233} [\href{https://arxiv.org/abs/2405.00112}{{\ttfamily
  2405.00112}}].

\bibitem{QROCODILE:2024nqm}
{\scshape QROCODILE} collaboration, \emph{{First Sub-MeV Dark Matter Search
  with the QROCODILE Experiment Using Superconducting Nanowire Single-Photon
  Detectors}}, \href{https://doi.org/10.1103/4hb6-f6jl}{\emph{Phys. Rev. Lett.}
  {\bfseries 135} (2025) 081002}
  [\href{https://arxiv.org/abs/2412.16279}{{\ttfamily 2412.16279}}].

\bibitem{Chen:2025cvl}
M.~Chen, V.~Takhistov, K.~Nakayama and K.~Hattori, \emph{{Light dark matter
  detection with sub-eV transition-edge sensors}},
  \href{https://doi.org/10.1103/74pr-jb4t}{\emph{Phys. Rev. D} {\bfseries 113}
  (2026) 036006} [\href{https://arxiv.org/abs/2506.10070}{{\ttfamily
  2506.10070}}].

\bibitem{Schwemmbauer:2025evp}
C.~Schwemmbauer et~al., \emph{{First direct search for light dark matter
  interactions in a transition-edge sensor}},
  \href{https://doi.org/10.1103/zytz-n25v}{\emph{Phys. Rev. D} {\bfseries 114}
  (2026) 012006} [\href{https://arxiv.org/abs/2506.18982}{{\ttfamily
  2506.18982}}].

\bibitem{Bardeen:1957mv}
J.~Bardeen, L.~N. Cooper and J.~R. Schrieffer, \emph{{Theory of
  superconductivity}},
  \href{https://doi.org/10.1103/PhysRev.108.1175}{\emph{Phys. Rev.} {\bfseries
  108} (1957) 1175}.

\bibitem{Lasenby:2021wsc}
R.~Lasenby and A.~Prabhu, \emph{{Dark matter{\textendash}electron scattering in
  materials: Sum rules and heterostructures}},
  \href{https://doi.org/10.1103/PhysRevD.105.095009}{\emph{Phys. Rev. D}
  {\bfseries 105} (2022) 095009}
  [\href{https://arxiv.org/abs/2110.01587}{{\ttfamily 2110.01587}}].

\bibitem{Lindhard:1954}
J.~Lindhard, \emph{{On the properties of a gas of charged particles}},
  {\emph{K. Dan. Vidensk. Selsk. Mat. Fys. Medd.} {\bfseries 28} (1954) 1}.

\end{thebibliography}\endgroup

\end{document}